\documentclass[12pt]{article}

\input{tcilatex}

\begin{document}

\bigskip

\bigskip\ 

\begin{center}
\textbf{TOWARDS AN ASHTEKAR FORMALISM IN EIGHT DIMENSIONS}

\bigskip\ 

\smallskip\ 

J. A. Nieto\footnote[1]{%
nieto@uas.uasnet.mx}

\smallskip\ 

\textit{Facultad de Ciencias F\'{\i}sico-Matem\'{a}ticas de la Universidad
Aut\'{o}noma}

\textit{de Sinaloa, 80010, Culiac\'{a}n Sinaloa, M\'{e}xico}

\bigskip\ 

\bigskip\ 

\textbf{Abstract}
\end{center}

We investigate the possibility of extending the Ashtekar theory to eight
dimensions. Our approach relies on two notions: the octonionic structure and
the MacDowell-Mansouri formalism generalized to a spacetime of signature $%
1+7 $. The key mathematical tool for our construction is the self-dual
(antiself-dual) four-rank fully antisymmetric octonionic tensor. Our results
may be of particular interest in connection with a possible formulation of $%
M-$theory via matroid theory.

\bigskip\ 

\bigskip\ 

\bigskip\ 

\bigskip\ 

\bigskip\ 

Pacs numbers: 04.60.-m, 04.65.+e, 11.15.-q, 11.30.Ly

February, 2005

\newpage \noindent \textbf{1. Introduction}

\bigskip

One of the most interesting candidate for the so called $M-$theory (see [1]
and Refs. therein) is a theory of a $2+2-$brane embedded in $2+10$
dimensional background target spacetime (see [2] and Refs. therein). This
version of $M-$theory evolved from the observation [3] that the complex
structure of $N=2$ strings requires a target spacetime of signature $2+2$
rather than $1+9$ as the usual $N=1$ strings. Thus, a natural step forward
was to consider the $N=(2,1)$ heterotic string [4]. In this scenario, it was
observed [5] that a consistent $N=(2,1)$ string should consider right-movers
`living' in $2+2$ dimensions and left-movers in $2+10$ dimensions, rather
than in $1+9$.

One of the interesting aspects of $2+2-$branes is that their dynamic lead to
self-dual self gravity coupled to self-dual supermatter in $2+10$ dimensions
[5]. Meanwhile, a realistic $1+3$ dimensional gravitational theory should
arise from the `compactification' of $1+7$ dimensions out of the full $2+10-$%
dimensional spacetime background. Thus, it is natural to expect that the
self-duality gravitational property of the full $2+10$ theory is broken in
self-duality gravity in $1+3$ and self-duality gravity in $1+7$ dimensions
[6]. But, it is well known that self-duality gravity in $1+3$ [7] can be
treated by means of the Ashtekar formalism (see [8] and Refs. therein).
Therefore, a natural question is to see whether the self-duality gravity in $%
1+7$ dimensions also admits an Ashtekar formulation.

Motivated by the above observations in this work we study the possibility of
extending the Ashtekar theory to $1+7$ dimensional spacetime. Our procedure
is based on two notions; octonionic structure [9]-[11] and
MacDowell-Mansouri formalism [12] (see also Ref. [13]). Specifically, the
central idea is to take the recourse of the properties of the four-rank
octonionic tensor [14] in order to derive the analogue in $1+7$ dimensions
of the self-dual gravitational action in $1+3$ dimensions as proposed in
reference [15]-[16], which is an extension of the Jacobson-Smolin-Samuel
action [7] of the Ashtekar formalism.

Our formalism may have some connection with the Nishino and Rajpoot [6]
theory of a consistent selfdual supergravity in eight dimensions, with
reduced holonomy $G_{2}$ and the topological quantum field theory of Baulieu
et. al. [17-18]. Moreover, since extensions of the MacDowell-Mansouri
procedure have played an essential part in the developing of different
supergravity theories one should expect that our formalism may help to have
a better understanding of such an eight dimensional gravitational theories.

By using the octonion structure, in section 2, we generalize the
MacDowell-Mansouri theory to $1+7-$dimensional spacetime. In section 3, we
show that this generalization admits a self-dual (antiself-dual) extension
which can be identified with an Ashtekar theory in eight dimensions. In
section 4, we outline a possible canonical quantization of our formalism and
we make some final comments about the importance that the present work may
have in connection \ with the relation between $M-$theoy and $M$(atroid)
theory.

\bigskip\ 

\noindent \textbf{2. Self-duality and antiself-duality in eight dimensions}

\bigskip

Let us start recalling some useful aspects of the octonionic structure [9].
The concept of octonions may arise from the Hurwitz theorem:

\textbf{Theorem (Hurwitz, 1898): }\textit{Every normed algebra with an
identity is isomorphic to one of following four algebras: the real numbers,
the complex numbers, the quaternions, and the Cayley (octonion) numbers.}

Let

\begin{equation}
e_{1},e_{2},...,e_{8},  \tag{1}
\end{equation}%
be a basis of the octonionic algebra $\mathcal{O}$, and let

\begin{equation}
A=A^{i}e_{i}  \tag{2}
\end{equation}%
be an octonionic vector $A$ $\epsilon $ $\mathcal{O}$ relative to this
basis, where the index $i$ runs from $1$ to $8$. Here, $%
A^{1},A^{2},...,A^{8}\epsilon R.$ Further, assume that the basis (1) is
orthonormal with bi-linear symmetric non-degenerate scalar product given by

\begin{equation}
<e_{i}\mid e_{j}>=\delta _{ij},  \tag{3}
\end{equation}%
where $\delta _{ij}$ is the Kronecker delta, with $\delta _{ij}=0$ if $i\neq
j$ and $\delta _{ij}=1$ if $i=j$.

Now, take the multiplication table between the basis elements $e_{i}$ as

\begin{equation}
e_{i}e_{j}=C_{ij}^{k}e_{k},  \tag{4}
\end{equation}%
Here, $C_{ij}^{k}$ are the so called structure constants of octonions.

According to the multiplication table (4) the product $AB=D$ for $A,B$ and $%
D\epsilon $ $\mathcal{O}$ is given by

\begin{equation}
A^{i}B^{j}C_{ij}^{k}=D^{k}.  \tag{5}
\end{equation}%
A normed algebra is an algebra in which the composition law

\begin{equation}
<AB\mid AB>=<A\mid A><B\mid B>  \tag{6}
\end{equation}%
holds for any $A$, $B$ $\epsilon $ $\mathcal{O}$. It can be shown that this
expression is equivalent to (see, for instance, section 3.1 of ref. 10)

\begin{equation}
<AB\mid CD>+<CB\mid AD>=2<A\mid C><B\mid D>,  \tag{7}
\end{equation}%
where $A$, $B,C,D$ $\epsilon $ $\mathcal{O}$. Choosing

\begin{equation}
A\rightarrow e_{i},B\rightarrow e_{j},C\rightarrow e_{m},D\rightarrow e_{n},
\tag{8}
\end{equation}%
we find that (7) leads to

\begin{equation}
<e_{i}e_{j}\mid e_{m}e_{n}>+<e_{m}e_{j}\mid e_{i}e_{n}>=2<e_{i}\mid
e_{m}><e_{j}\mid e_{n}>.  \tag{9}
\end{equation}

Using (3) and (4), from (9) we obtain the key formula

\begin{equation}
C_{ij}^{k}C_{mn}^{l}\delta _{kl}+C_{mj}^{k}C_{in}^{l}\delta _{kl}=2\delta
_{im}\delta _{jn}.  \tag{10}
\end{equation}%
Let $e_{8}$ be the identity of the algebra $\mathcal{O}$. It can be shown
that (10) is reduced to

\begin{equation}
C_{ab}^{e}C_{cd}^{f}\delta _{ef}+C_{cb}^{e}C_{ad}^{f}\delta _{ef}=2\delta
_{ac}\delta _{bd}-\delta _{ab}\delta _{cd}-\delta _{cb}\delta _{ad}, 
\tag{11}
\end{equation}%
where the indices $a,b$ run from $1$ to $7$ and $C_{abc}$ is a completely
antisymmetric quantity. Moreover, it can be shown that the Cartan-Shouten's
equations [19] (used by Gursey and Tze [20] in a consistent octonionic
compactification of $D=11$ supergravity) can be derived from (11) (see Ref.
[21) details).

Let us write (11) in the form

\begin{equation}
C_{ab}^{e}C_{cd}^{f}\delta _{ef}-(\delta _{ac}\delta _{bd}-\delta
_{ad}\delta _{cb})+C_{cb}^{e}C_{ad}^{f}\delta _{ef}-(\delta _{ac}\delta
_{bd}-\delta _{ab}\delta _{cd})=0.  \tag{12}
\end{equation}%
This expression suggests to define

\begin{equation}
F_{abcd}\equiv C_{ab}^{e}C_{cd}^{f}\delta _{ef}-(\delta _{ac}\delta
_{bd}-\delta _{ad}\delta _{cb}).  \tag{13}
\end{equation}%
Therefore (12) gives

\begin{equation}
F_{abcd}+F_{cbad}=0.  \tag{14}
\end{equation}%
Thus, considering that $C_{ab}^{e}$ is completely antisymmetric object, from
(13) and (14) we discover that $F_{abcd}$ is also completely antisymmetric
tensor. Using this important cyclic property for $F_{abcd}$ it is not
difficult to show that

\begin{equation}
F_{abcd}=C_{e[ab}C_{c]d}^{e}.  \tag{15}
\end{equation}%
Substituting this result into (13) leads us to an expression that can be
connected with the parallelizability of seven sphere $S^{7}$. In this case
the object $C_{ab}^{e}$ can be identified with the `undressed' torsion
associated with $S^{7}$ (see Ref. [21] for details).

An alternative, but equivalent, definition of $F_{abcd}$ can be given by
means of the expression

\begin{equation}
F_{abcd}=\frac{1}{3!}\varepsilon _{abcdefg}C^{efg},  \tag{16}
\end{equation}%
where $\varepsilon _{abcdefg}$ is the completely antisymmetric symbol in
seven dimensions.

The duality expression in (16) can be extended to eight dimensions by
defining

\begin{equation}
\eta _{0abc}\equiv iC_{abc}  \tag{17}
\end{equation}%
and

\begin{equation}
\eta _{abcd}\equiv F_{abcd}.  \tag{18}
\end{equation}%
In fact the duality relation (22) can now be written as

\begin{equation}
\eta _{\mu \nu \alpha \beta }=\frac{i}{4!}\varepsilon _{\mu \nu \alpha \beta
\lambda \rho \sigma \tau }\eta ^{\lambda \rho \sigma \tau }.  \tag{19}
\end{equation}%
Here, the indices $\mu ,\nu $ run from $0$ to $7.$ The formula (19) means
that $\eta _{\mu \nu \alpha \beta }$ is self-dual (or antiself-dual putting $%
-i$ instead of $i$ in (19)). In the expression (19) we used a Minkowski
metric $\eta _{\mu \nu }$ with signature $1+7$ and this is the reason why we
introduced a complex structure in (17)$.$

Using (11),(13), (17) and (18) one can verify that four-rank completely
antisymmetric tensor $\eta _{\mu \nu \alpha \beta }$ satisfies the relations
[14],

\begin{equation}
\eta _{\mu \nu \alpha \beta }\eta ^{\tau \sigma \alpha \beta }=6\delta _{\mu
\nu }^{\tau \sigma }+4\eta _{\mu \nu }^{\tau \sigma },  \tag{20}
\end{equation}%
\begin{equation}
\eta _{\mu \nu \alpha \beta }\eta ^{\tau \nu \alpha \beta }=42\delta _{\mu
}^{\tau },  \tag{21}
\end{equation}%
\begin{equation}
\eta _{\mu \nu \alpha \beta }\eta ^{\mu \nu \alpha \beta }=336,  \tag{22}
\end{equation}%
where $\delta _{\mu }^{\tau }$ is the Kronecker delta and $\delta _{\mu \nu
}^{\tau \sigma }=\delta _{\mu }^{\tau }\delta _{\nu }^{\sigma }-\delta _{\mu
}^{\sigma }\delta _{\nu }^{\tau }$.

Let us define the dual of a Yang-Mills field strength $F_{\mu \nu }^{i}$ by
(see Refs. in [22])

\begin{equation}
^{\ast }F_{\mu \nu }^{i}=\frac{1}{2}\eta _{\mu \nu }^{\alpha \beta
}F_{\alpha \beta }^{i}.  \tag{23}
\end{equation}%
Using this definition we now write the self-dual part of $F_{\mu \nu }^{i}$
as

\begin{equation}
^{+}F_{\mu \nu }^{i}=\frac{1}{2}(F_{\mu \nu }^{i}+^{\ast }F_{\mu \nu }^{i}).
\tag{24}
\end{equation}

Observe that

\begin{equation}
^{\ast \ast }F_{\mu \nu }^{i}=3F_{\mu \nu }^{i}+2^{\ast }F_{\mu \nu }^{i}, 
\tag{25}
\end{equation}%
and therefore

\begin{equation}
^{\ast +}F_{\mu \nu }^{i}=3^{+}F_{\mu \nu }^{i},  \tag{26}
\end{equation}%
where we used the relation (20). Thus, except for a numerical factor $%
^{+}F_{\mu \nu }^{i}$ satifies the usual self-dual relation.

\bigskip\ 

\noindent \textbf{3. MacDowell-Mansouri in eight dimensions}

\bigskip

Let us consider the MacDowell-Mansouri type action in a $2+10-$dimensional
spacetime

\begin{equation}
S=\int_{M^{2+10}}\Omega ^{\hat{\mu}\hat{\nu}\hat{\alpha}\hat{\beta}}\mathcal{%
R}_{\hat{\mu}\hat{\nu}}^{\hat{A}\hat{B}}\mathcal{R}_{\hat{\alpha}\hat{\beta}%
}^{\hat{C}\hat{D}}\Omega _{\hat{A}\hat{B}\hat{C}\hat{D}}.  \tag{27}
\end{equation}%
Here, $M^{2+10}$ is a smooth spacetime manifold with an associated $2+10-$%
signature. Further, $\Omega ^{\hat{\mu}\hat{\nu}\hat{\alpha}\hat{\beta}}$ is
an arbitrary function in $M^{10+2}$ and $\Omega _{ABCD}$ is a $SO(2,10)$
group invariant quantity. Furthermore, $\mathcal{R}_{\hat{\mu}\hat{\nu}}^{%
\hat{A}\hat{B}}$ is given by

\begin{equation}
\mathcal{R}_{\hat{\mu}\hat{\nu}}^{\hat{A}\hat{B}}=R_{\hat{\mu}\hat{\nu}}^{%
\hat{A}\hat{B}}+\Sigma _{\hat{\mu}\hat{\nu},}^{\hat{A}\hat{B}}  \tag{28}
\end{equation}%
with\ 

\begin{equation}
R_{\hat{\mu}\hat{\nu}}^{\hat{A}\hat{B}}=\partial _{\hat{\mu}}\omega _{\hat{%
\nu}}^{\hat{A}\hat{B}}-\partial _{\hat{\nu}}\omega _{\hat{\mu}}^{\hat{A}\hat{%
B}}+\omega _{\hat{\mu}}^{\hat{A}\hat{C}}\omega _{\hat{\nu}\hat{C}}^{\hat{B}%
}-\omega _{\hat{\mu}}^{\hat{B}\hat{C}}\omega _{\hat{\nu}\hat{C}}^{\hat{A}} 
\tag{29}
\end{equation}%
and

\begin{equation}
\Sigma _{\hat{\mu}\hat{\nu}}^{\hat{A}\hat{B}}=e_{\hat{\mu}}^{\hat{A}}e_{\hat{%
\nu}}^{\hat{B}}-e_{\hat{\mu}}^{\hat{B}}e_{\hat{\nu}}^{\hat{A}}.  \tag{30}
\end{equation}

We shall assume that the spacetime manifold $M^{2+10}$ can be broken up in
the form $M^{10+2}$=$M^{1+3}\times M^{1+7}$ and that $\Omega ^{\hat{\mu}\hat{%
\nu}\hat{\alpha}\hat{\beta}}$ and $\Omega _{\hat{A}\hat{B}\hat{C}\hat{D}}$
can be chosen in such a way that (27) takes the form

\begin{equation}
S=\int_{M^{1+3}}\varepsilon ^{ijkl}\mathcal{R}_{ij}^{AB}\mathcal{R}%
_{kl}^{CD}\varepsilon _{ABCD}+\int_{M^{1+7}}\eta ^{\mu \nu \alpha \beta }%
\mathcal{R}_{\mu \nu }^{\hat{a}\hat{b}}\mathcal{R}_{\alpha \beta }^{\hat{c}%
\hat{d}}\eta _{\hat{a}\hat{b}\hat{c}\hat{d}}.  \tag{31}
\end{equation}%
Here, $\mathcal{R}_{ij}^{AB}$ and $\mathcal{R}_{\mu \nu }^{\hat{a}\hat{b}}$
are a reduced version of (28), namely

\begin{equation}
\mathcal{R}_{ij}^{AB}=R_{ij}^{AB}+\Sigma _{ij}^{AB}  \tag{32}
\end{equation}%
and

\begin{equation}
\mathcal{R}_{\mu \nu }^{\hat{a}\hat{b}}=R_{\mu \nu }^{\hat{a}\hat{b}}+\Sigma
_{\mu \nu }^{\hat{a}\hat{b}},  \tag{33}
\end{equation}%
with the corresponding definitions (29) and (30) for $R_{ij}^{AB},\Sigma
_{ij}^{AB},R_{\mu \nu }^{\hat{a}\hat{b}}$ and $\Sigma _{\mu \nu }^{\hat{a}%
\hat{b}}$. In addition, we assume that $\mathcal{R}_{ij}^{AB}$ and $\mathcal{%
R}_{\mu \nu }^{\hat{a}\hat{b}}$ `live' in $M^{1+3}$ and $M^{1+7}$
respectively.

We recognize in the first term in the action (31) the ordinary
MacDowell-Mansouri action. Hence, let us focus only in the second term of
the action (31). Using, (33) such a term becomes

\begin{equation}
\begin{array}{c}
S_{1+7}\equiv \int_{M^{1+7}}\eta ^{\mu \nu \alpha \beta }\mathcal{R}_{\mu
\nu }^{\hat{a}\hat{b}}\mathcal{R}_{\alpha \beta }^{\hat{c}\hat{d}}\eta _{%
\hat{a}\hat{b}\hat{c}\hat{d}}=\int_{M^{1+7}}\eta ^{\mu \nu \alpha \beta
}R_{\mu \nu }^{\hat{a}\hat{b}}R_{\alpha \beta }^{\hat{c}\hat{d}}\eta _{\hat{a%
}\hat{b}\hat{c}\hat{d}} \\ 
\\ 
+2\int_{M^{1+7}}\eta ^{\mu \nu \alpha \beta }\Sigma _{\mu \nu }^{\hat{a}\hat{%
b}}R_{\alpha \beta }^{\hat{c}\hat{d}}\eta _{\hat{a}\hat{b}\hat{c}\hat{d}%
}+\int_{M^{1+7}}\eta ^{\mu \nu \alpha \beta }\Sigma _{\mu \nu }^{\hat{a}\hat{%
b}}\Sigma _{\alpha \beta }^{\hat{c}\hat{d}}\eta _{\hat{a}\hat{b}\hat{c}\hat{d%
}}.%
\end{array}
\tag{34}
\end{equation}%
Presumably, the first term is a total derivative, that is a topological
invariant. In fact, this term has the same form that the Euler
characteristic in four dimensions which is topological invariant. However,
to our knowledge such a term has not been considered in the mathematical
context. Thus, such a term may be the source of some interest in topology.

The second term in (34) leads to a type Einstein-Hilbert action in eight
dimensions plus an additional interacting term of the form

\begin{equation}
\sim \int_{M^{1+7}}\eta ^{\mu \nu \alpha \beta }R_{\mu \nu \alpha \beta }, 
\tag{35}
\end{equation}%
which should vanish after using the algebraic Bianchi identities for $R_{\mu
\nu \alpha \beta }.$ Finally, the last term in (34) may be interpreted as a
cosmological constant term in eight dimensions. In fact, it is interesting
to observe that the combination

\begin{equation}
\eta ^{\mu \nu \alpha \beta }\Sigma _{\mu \nu }^{\hat{a}\hat{b}}\Sigma
_{\alpha \beta }^{\hat{c}\hat{d}}\eta _{\hat{a}\hat{b}\hat{c}\hat{d}}=4\eta
^{\mu \nu \alpha \beta }e_{\mu }^{\hat{a}}e_{\nu }^{\hat{b}}e_{\alpha }^{%
\hat{c}}e_{\beta }^{\hat{d}}\eta _{\hat{a}\hat{b}\hat{c}\hat{d}}  \tag{36}
\end{equation}%
does not correspond to the expression $det(e_{\mu }^{\hat{a}})$ given by

\begin{equation}
det(e_{\mu }^{\hat{a}})=\frac{1}{8!}\varepsilon ^{\mu _{0}...\mu
_{7}}\varepsilon _{\hat{a}_{0}...\hat{a}_{7}}e_{\mu _{0}}^{\hat{a}%
_{0}}...e_{\mu _{7}}^{\hat{a}_{7}},  \tag{37}
\end{equation}%
which gives the usual cosmological constant term. Nevertheless, we have
shown that up to topological invariant the action $S_{1+7}$ has a similar
form as the Einstein-Hilbert action with cosmological constant in a $1+7$
dimensional spacetime.

\bigskip\ 

\noindent \textbf{4. Ashtekar formalism in }$1+7-$\textbf{dimensional
spacetime}

\bigskip

Let us start by introducing the dual (antidual) of $\mathcal{R}_{\mu \nu }^{%
\hat{a}\hat{b}}$ in the form

\begin{equation}
^{\star }\mathcal{R}_{\mu \nu }^{\hat{a}\hat{b}}=\frac{1}{2}\eta _{\hat{c}%
\hat{d}}^{\hat{a}\hat{b}}\mathcal{R}_{\mu \nu }^{\hat{c}\hat{d}}.  \tag{38}
\end{equation}%
Observe that $^{\star }\mathcal{R}_{\mu \nu }^{\hat{a}\hat{b}}$ is dual of $%
\mathcal{R}_{\mu \nu }^{\hat{a}\hat{b}}$ in the internal indices $\hat{a}$
and $\hat{b}.$ We shall define the self-dual (anti self-dual) parts $^{\pm }%
\mathcal{R}_{\mu \nu }^{\hat{a}\hat{b}}$ of $\mathcal{R}_{\mu \nu }^{\hat{a}%
\hat{b}}$ in the form%
\begin{equation}
^{+}\mathcal{R}_{\mu \nu }^{\hat{a}\hat{b}}=\frac{1}{2}(\mathcal{R}_{\mu \nu
}^{\hat{a}\hat{b}}+^{\ast }\mathcal{R}_{\mu \nu }^{\hat{a}\hat{b}}). 
\tag{39}
\end{equation}%
Since

\begin{equation}
^{\star \star }\mathcal{R}_{\mu \nu }^{\hat{a}\hat{b}}=3\mathcal{R}_{\mu \nu
}^{\hat{a}\hat{b}}+2^{\star }\mathcal{R}_{\mu \nu }^{\hat{a}\hat{b}}, 
\tag{40}
\end{equation}%
we find

\begin{equation}
^{\star +}\mathcal{R}_{\mu \nu }^{\hat{a}\hat{b}}=3^{+}\mathcal{R}_{\mu \nu
}^{\hat{a}\hat{b}}.  \tag{41}
\end{equation}%
Thus, up to a numerical factor we see that $^{+}\mathcal{R}_{\mu \nu }^{\hat{%
a}\hat{b}}$ plays, in fact, the role of the self-dual part of $\mathcal{R}%
_{\mu \nu }^{\hat{a}\hat{b}}.$ It turns out to be convenient to write (39) as

\begin{equation}
^{\pm }\mathcal{R}_{\mu \nu }^{\hat{a}\hat{b}}=\frac{1}{2}^{\pm }B_{\hat{c}%
\hat{d}}^{\hat{a}\hat{b}}\mathcal{R}_{\mu \nu }^{\hat{c}\hat{d}},  \tag{42}
\end{equation}%
where

\begin{equation}
^{\pm }B_{\hat{c}\hat{d}}^{\hat{a}\hat{b}}=\frac{1}{2}(\delta _{\hat{c}\hat{d%
}}^{\hat{a}\hat{b}}\pm \eta _{\hat{c}\hat{d}}^{\hat{a}\hat{b}}).  \tag{43}
\end{equation}%
Here, we include the aniself-dual part of $\mathcal{R}_{\mu \nu }^{\hat{a}%
\hat{b}}$ for a possible generalization.

Now, we would like to propose the action

\begin{equation}
^{\pm }\mathcal{S}_{1+7}=\frac{1}{^{+}\tau }\int_{M^{1+7}}\eta ^{\mu \nu
\alpha \beta +}\mathcal{R}_{\mu \nu }^{\hat{a}\hat{b}+}\mathcal{R}_{\alpha
\beta }^{\hat{c}\hat{d}}\eta _{\hat{a}\hat{b}\hat{c}\hat{d}}+\frac{1}{%
^{-}\tau }\int_{M^{1+7}}\eta ^{\mu \nu \alpha \beta -}\mathcal{R}_{\mu \nu
}^{\hat{a}\hat{b}-}\mathcal{R}_{\alpha \beta }^{\hat{c}\hat{d}}\eta _{\hat{a}%
\hat{b}\hat{c}\hat{d}},  \tag{44}
\end{equation}%
which is a generalization of the action (34). Here, $^{+}\tau $ and $%
^{-}\tau $ are two constant parameters. Although our main goal is to show
that the first term or the second term of the action (44) lead to an
Ashtekar formalism in a $1+7-$dimensional spacetime, it is interesting to
mention that, in principle, following the formalism of the Ref. [23], the
action (44) may be taken as the starting point in an $S-$duality
consideration of gravity in eight dimensions.

Let us focus in the self-dual part of (44):

\begin{equation}
^{+}\mathcal{S}_{1+7}=\frac{1}{^{+}\tau }\int_{M^{1+7}}\eta ^{\mu \nu \alpha
\beta +}\mathcal{R}_{\mu \nu }^{\hat{a}\hat{b}+}\mathcal{R}_{\alpha \beta }^{%
\hat{c}\hat{d}}\eta _{\hat{a}\hat{b}\hat{c}\hat{d}}.  \tag{45}
\end{equation}%
Presumably, most of the computations that we shall develop below in
connection with $^{+}\mathcal{S}_{1+7}$ may also be applied to the
antiselfdual sector $^{-}\mathcal{S}_{1+7}.$ It is worth mentioning that the
action (45) is the analogue of the action proposed by Nieto \textit{et al}
[15]-[16] in four dimensions. Let us start observing that since

\begin{equation}
^{+}\mathcal{R}_{\mu \nu }^{\hat{a}\hat{b}}=^{+}R_{\mu \nu }^{\hat{a}\hat{b}%
}+^{+}\Sigma _{\mu \nu }^{\hat{a}\hat{b}},  \tag{46}
\end{equation}%
one finds that the action (45) becomes

\begin{equation}
^{+}\mathcal{S}_{1+7}=\frac{1}{^{+}\tau }\int_{M^{1+7}}(T+K+C),  \tag{47}
\end{equation}%
with

\begin{equation}
T=\eta ^{\mu \nu \alpha \beta +}R_{\mu \nu }^{\hat{a}\hat{b}+}R_{\alpha
\beta }^{\hat{c}\hat{d}}\eta _{\hat{a}\hat{b}\hat{c}\hat{d}},  \tag{48}
\end{equation}

\begin{equation}
K=2\eta ^{\mu \nu \alpha \beta +}\Sigma _{\mu \nu }^{\hat{a}\hat{b}%
+}R_{\alpha \beta }^{\hat{c}\hat{d}}\eta _{\hat{a}\hat{b}\hat{c}\hat{d}}, 
\tag{49}
\end{equation}%
and

\begin{equation}
C=\eta ^{\mu \nu \alpha \beta +}\Sigma _{\mu \nu }^{\hat{a}\hat{b}+}\Sigma
_{\alpha \beta }^{\hat{c}\hat{d}}\eta _{\hat{a}\hat{b}\hat{c}\hat{d}}. 
\tag{50}
\end{equation}%
In order to clarify the meaning of the different terms of the action (47) it
is convenient to compute the expression $^{+}B_{\mu \nu }^{\hat{a}\hat{b}%
+}B_{\alpha \beta }^{\hat{c}\hat{d}}\eta _{\hat{a}\hat{b}\hat{c}\hat{d}}$
where $^{+}B_{\mu \nu }^{\hat{a}\hat{b}}$ is given by (43). Using (20) we
find

\begin{equation}
\begin{array}{c}
^{+}B_{\hat{e}\hat{f}}^{\hat{a}\hat{b}+}B_{\hat{g}\hat{h}}^{\hat{c}\hat{d}%
}\eta _{\hat{a}\hat{b}\hat{c}\hat{d}}=\frac{1}{4}(\delta _{\hat{e}\hat{f}}^{%
\hat{a}\hat{b}}+\eta _{\hat{e}\hat{f}}^{\hat{a}\hat{b}})(\delta _{\hat{g}%
\hat{h}}^{\hat{c}\hat{d}}+\eta _{\hat{g}\hat{h}}^{\hat{c}\hat{d}})\eta _{%
\hat{a}\hat{b}\hat{c}\hat{d}} \\ 
\\ 
=12(\delta _{\hat{e}\hat{f}\hat{g}\hat{h}}+\eta _{\hat{e}\hat{f}\hat{g}\hat{h%
}}).%
\end{array}
\tag{51}
\end{equation}%
Using (51) we discover that $T,K$ and $C$ give

\begin{equation}
T=12(\eta ^{\mu \nu \alpha \beta }R_{\mu \nu }^{\hat{a}\hat{b}}R_{\alpha
\beta }^{\hat{c}\hat{d}}\eta _{\hat{a}\hat{b}\hat{c}\hat{d}}+\eta ^{\mu \nu
\alpha \beta }R_{\mu \nu }^{\hat{a}\hat{b}}R_{\alpha \beta }^{\hat{c}\hat{d}%
}\delta _{\hat{a}\hat{b}\hat{c}\hat{d}}),  \tag{52}
\end{equation}

\begin{equation}
K=24(\eta ^{\mu \nu \alpha \beta }\Sigma _{\mu \nu }^{\hat{a}\hat{b}%
}R_{\alpha \beta }^{\hat{c}\hat{d}}\eta _{\hat{a}\hat{b}\hat{c}\hat{d}}+\eta
^{\mu \nu \alpha \beta }\Sigma _{\mu \nu }^{\hat{a}\hat{b}}R_{\alpha \beta
}^{\hat{c}\hat{d}}\delta _{\hat{a}\hat{b}\hat{c}\hat{d}}),  \tag{53}
\end{equation}%
and

\begin{equation}
C=12(\eta ^{\mu \nu \alpha \beta }\Sigma _{\mu \nu }^{\hat{a}\hat{b}}\Sigma
_{\alpha \beta }^{\hat{c}\hat{d}}\eta _{\hat{a}\hat{b}\hat{c}\hat{d}}+\eta
^{\mu \nu \alpha \beta }\Sigma _{\mu \nu }^{\hat{a}\hat{b}}\Sigma _{\alpha
\beta }^{\hat{c}\hat{d}}\delta _{\hat{a}\hat{b}\hat{c}\hat{d}}).  \tag{54}
\end{equation}%
The first and the second term in $T$ can be interpreted as the analogue of
the Euler and Pontrjagin topological invariants in $1+3$ dimensions
respectively. The first term in $K$ corresponds to the second term in (34),
while the second term in (53) is of the type (35). Similarly, the first term
in $C$ corresponds to the third term in (34), while due to the antisymmetric
properties of $\eta ^{\mu \nu \alpha \beta }$ one finds that the second term
of $C$ is identically zero. This shows the action (45) is a generalization
of the action (34). In fact, up to the Pontrjagin topological invariant and
additional contribution of the type (35) the action (45) corresponds exactly
to the action (34).

We would like to show now that the $K$ term admits an Ashtekar
interpretation in $1+7$ dimensional spacetime. Using (20), it is not
difficult to see that $K$ can also be written as

\begin{equation}
K=6\eta ^{\mu \nu \alpha \beta }e_{\mu }^{\hat{a}}e_{\nu }^{\hat{b}%
+}R_{\alpha \beta }^{\hat{c}\hat{d}}\eta _{\hat{a}\hat{b}\hat{c}\hat{d}%
}+12\eta ^{\mu \nu \alpha \beta +}R_{\mu \nu \alpha \beta }.  \tag{55}
\end{equation}%
We recognize in the first term of (55) the typical form of the integrand of
the Jacobson-Smolin-Samuel's action [7] of the Ashtekar formalism . The last
term of (55) is of the type of (35) but with $R_{\mu \nu \alpha \beta }$
replaced by $^{+}R_{\mu \nu \alpha \beta }.$ Indeed, such last term appears
because we insisted on writing the first term \textit{a la}
Jacobson-Smolin-Samuel's form, otherwise the expression (53) represented the
final result. Nevertheless, according to the experience gained in four
dimensions [8] one should expect that using (55) rather than (53) some
advantage at the level of canonical quantization should be present.

\bigskip\ 

\noindent \textbf{5. Final comments}

\bigskip

Let us focus in the $K$ sector of the action (47):

\begin{equation}
^{+}\mathcal{S}_{1+7}^{K}=\frac{1}{^{+}\tau }\int_{M^{1+7}}K,  \tag{56}
\end{equation}%
with $K$ given by (55). Following the experience in $1+3-$dimensional
spacetime, the steps towards a canonical quantization of the action $^{+}%
\mathcal{S}_{1+7}^{K}$ should be straightforward. Let us outline the main \
ideas. One first, introduces a foliation of the spacetime $M^{1+7}$ by a
family of space-like hypersurface labelled by $t=const..$\textit{\ }The
foliation is characterized by the decomposition: $t^{\mu }=Nn^{\mu }+N^{\mu }
$, where $t^{\mu }$ is a real vector field whose integral curves intersect
each leaf of the foliation precisely once, $N$ is the lapse function and $%
N^{\mu }$ is the shift function. Here, $n^{\mu }$ is a unit normal for the
leaves of the foliation, with $N^{\mu }n_{\mu }=0$. The strategy now is to
decompose the various fields on the action $^{+}\mathcal{S}_{1+7}^{K}$ by
means of the $n^{\mu }$ and the projector $q_{\nu }^{\mu }=\delta _{\nu
}^{\mu }+n^{\mu }n_{\nu }.$ At the end, it should be possible to write the
lagrangian $K$ in the form $p\dot{q}-H$, where $H$, the extended
Hamiltonian, contains the constraints associated with $N,N^{\mu }$ and the
connection $^{+}\omega _{\mu }^{\hat{a}\hat{b}}$. The quantization of the
constraints may be carried out by using the Dirac's procedure for
quantization of constrained systems. However, one should expect that this
quantization program goes beyond not only of the Dirac formalism but of the
four dimensional case. For instance, the problem for finding the correct
inner product in the space of physical states may be as difficult as in four
dimensions. Another difficult problem that may bring new interesting results
is the `loop' representation of quantum gravity in eight dimensions. In the
case of four dimensions the loop representation is related to different
properties of knot theory which is intrinsically three dimensional.
Therefore, a `loop' representation of quantum gravity in eight dimension
should necessarily require to go beyond knot theory.

Summarizing, using a octonionic structure and a MacDowell-Mansouri approach
in this work we have shown that an Ashtekar formalism makes sense in a $1+7$
dimensional spacetime. The proposed action (45) presents the analogue of the
terms developed in the Ref. [15]-[16]. In particular, we showed that the
action (45) leads to the Jacobson-Smolin-Samuel's action generalized to
eight dimensions. A classical canonical quantization of the action (45)
seems to be straightforward. However, its quantization may present
interesting new problems beyond the four dimensional case such as the inner
product in the space of physical states and the `loop' representation of
quantum gravity in $1+7$ dimensional spacetime.

Our discussion was focused on the self-dual sector of the action (44). But
what about the antiself-dual sector? In the case of $1+3$ dimensional
spacetime, as Jacobson and Smolin emphasized [7], the decomposition of the
self-dual and antiself-dual sectors is possible becuase of the Lie algebra
property $so(1,3)=su(2)\oplus su(2)$. In $1+7$-dimensional spacetime the
situation is not so simple since the closest decompostion to the case $%
so(1,3)$ of the Lie algebra $so(1,7)$ associated with $SO(1,7)$ seems to be $%
so(1,7)=g_{2}\oplus L_{\func{Im}(\mathcal{O})}\oplus R_{\func{Im}(\mathcal{O}%
)}$, where $g_{2}$ is the Lie algebra of the exceptional group $G_{2}$, $L_{%
\func{Im}(\mathcal{O})}$ ($R_{\func{Im}(\mathcal{O})})$ is the space of
linear transformations of $\mathcal{O}$ given by left (right) multiplication
by imaginary octonions. To see that this is really the case observe that $%
\dim so(1,7)=28$ and $\dim g_{2}+\dim L_{\func{Im}(\mathcal{O})}+\dim R_{%
\func{Im}(\mathcal{O})}=14+7+7$ (see Ref. [11] for details)$.$ Thus while
the self-dual sector is related to the exceptional group $G_{2}$ one finds
the intriguing result that the antiself-dual sector should be related to two
copies of $\func{Im}(\mathcal{O})$, the $7-$dimensional space consisting of
all imaginary octonions. In some sense the $1+7$ case seems to be more
interesting than the $1+3$-dimensional case beacuse it is linked to the
exceptional group $G_{2}$ which, as it is known, leads to a reduced holonomy
of a $7-$dimensional compactification of $D=11$ supergravity (see [6] and
Refs. therein).

By using a self-dual 2-form as a primary variable Capovilla et al [24]
connected the Jacobson-Smolin-Samuel action with the Plebanski action of
complex general relativity [25]. This connection is important because
establishes that the Ashtekar canonical formalism corresponds to the
(3+1)-decomposition of the first order formalism of the Plebanski action. It
turns out that the Plebanki's idea has been extended to higher dimensions
(see [26]). Thus, it seems interesting for further research to establish a
connection between our work and such a higher dimensional extension of the
Plebanki action.

Beyond the Ashtekar formalism in eight dimensions, the present work may
appear interesting in connection with the proposal [27]-[32] that oriented
matroid theory [33] may be the underlaying mathematical theory supporting $M-
$theory. The reason for this is that the four-rank completely antisymmetric
tensor $\eta _{\mu \nu \alpha \beta }$ is closely linked to the chirotope $%
\chi _{\mu \nu \alpha \beta }$ associated to the uniform matroid $U_{4,8}$
as it was shown in references [28]-[30]. Because the Fano matroid is not
orientable [33], one finds that $\eta _{\mu \nu \alpha \beta }$ is not a
chirotope, but this may suggest to use a four-rank chirotopes instead of $%
\eta _{\mu \nu \alpha \beta }$ in the action (45). In the case of four
dimensional spacetime one uses in analogue action as (45) the completely
antisymmetric symbol $\varepsilon _{\mu \nu \alpha \beta }$ which is a
chirotope associated with an underlaying uniform matroid $U_{4,4}$.
Therefore, beyond the tensor $\eta _{\mu \nu \alpha \beta }$ oriented
matroid theory offers the possibility of using a chirotope of rank four in a
gravitational studies in eight dimensions.

\bigskip\ 

\noindent \textbf{Acknowledgment:} I would like to thank H. Garc\'{\i}a-Compe%
\'{a}n C. Ram\'{\i}rez and O. Obreg\'{o}n for helpful comments.

\end{document}